# In-situ temperature and major species measurements of sooting flames based on short-gated spontaneous Raman scattering

Hu Meng[1], Yihua Ren[1]*, Heinz Pitsch[1]

[1] Institute for Combustion Technology, RWTH Aachen University, Templergraben 64, Aachen 52056, Germany

∗ Corresponding author. E-mail address: y.ren@itv.rwth-aachen.de (Y. Ren).

# Abstract

Spontaneous Raman spectroscopy (SRS) is a conventional *in-situ* laser diagnostic method that has been widely used for measurements of temperature and major species. However, SRS in sooting flames suffers from strong interference including laser-induced fluorescence, laser-induced incandescence, and flame luminosity, which is a long-lasting challenge. This work introduces a low-cost, easy-to-implement, and calibration-free SRS thermometry in sooting flames based on a 355-nm nanosecond laser beam. Several strategies were utilized to increase the signal-to-noise ratio and suppress the interference: (1) nanosecond ICCD gate width; (2) optimized ICCD gate delay; (3) specially designed focusing shape of the laser beam; (4) ultraviolet polarizer filter. The temperature was obtained by fitting the contour of Stokes-Raman spectra of $N_2$ molecules, which does not require additional calibration. Based on the measured temperature, the mole fraction of major species can be obtained with calibration. This method was used in the temperature and major species measurements of an ethylene-based counterflow diffusion flame. The experimental results show an excellent agreement with the simulation results, demonstrating the feasibility of performing non-intrusive laser diagnostics of sooting and other particle-laden flames accurately.

# 1. Introduction

Spontaneous Raman spectroscopy (SRS) is a widely-used *in-situ* laser diagnostic technique for combustion and other gas-phase reactive flows. This non-contact, simultaneous measurement technique is perfectly suitable for the high-temperature and high-pressure combustion environment. The flame temperature and the mole fraction of major species can be obtained from the Raman spectra. However, the Raman spectroscopy is limited by its weak signal and thus its high sensitivity to interference signals. Considering that the typical Stokes Raman spectra are on the order of $10^{-3}$ of laser-induced Rayleigh scattering [1], the SRS signals are always overlapped by different interference, e.g., laser-induced fluorescence (LIF) by polycyclic aromatic hydrocarbon (PAHs), laser-induced incandescence (LII) by soot particles, and flame luminosity. The stokes-Raman scattering can be obtained when the flame is clean, non-sooting, and stable.

Many methods have been developed to isolate SRS from the interference signals based on the different properties of the SRS [2]. The first approach is to select the excitation laser wavelength for which the SRS is maximized while LIF and LII signals are minimized. The Raman scattering intensity depends on the laser frequency of the fourth power ($Raman \propto 1/\lambda_{laser}^4$). Thus, the shorter the laser wavelength, the stronger the SRS intensity [3]. Furthermore, a UV laser source induces a stronger LIF signal, while the strong LII signals are usually induced by the IR laser source [4]. Rabenstein and Leipertz [5] found that the laser with 355 nm wavelength was particularly suitable for SRS in hydrocarbon sooting flame. Dreyer et al. [6] investigated the SRS in $C_3H_8$/air flames at 532 and 355 nm excitation wavelength, indicating that 355 nm has a comparable signal-to-noise ratio (SNR) to 532 nm for a sooting flame. Egermann et al. [7] concluded that for a heavy soot loading flame, the interference signals at the laser wavelength of 266 nm are weaker than 355 nm. Taking all factors into account, the 355-nm laser was chosen for the measurement of sooting flames in the present work.

The second approach is to make use of the polarization of the Raman scattering.

Considering that the LIF and LII signals are depolarized due to the collision between molecules in the emission process [8], the clean Raman scattering signal can be obtained by two sequential superimposed laser beams with different polarizations [9,10] or using two imaging systems with a polarization filter with different directions [6,7,11–13]. Nevertheless, this technique has limitations because the LIF/LII signals are not 100% depolarized, which means some LIF/LII signals remain in the processed Raman signal. Furthermore, the data post-processing needs two measurements to get one SRS profile.

The third approach is to utilize the different lifetime between Raman scattering and interference. The lifetime of Rayleigh/Raman scattering is approximate $10^{-12}$ seconds or less, which is much smaller than LIF ($10^{-10}$ - $10^{-5}$ s) and LII ($\sim 10^{-7}$ s) [3]. A Kerr gate can be used to separate the Raman scattering from the collected signals [14–16]. Another approach is to use a picosecond-gated camera or a photomultiplier tube [17–19]. These time-domain methods need the picosecond-level pulse laser, which is usually expensive and not easy to implement.

Besides those approaches to suppress interference signals, many strategies have also been developed to maximize the SRS intensity. The most direct way is by maximizing the laser intensity. However, a laser-induced breakdown will occur when the laser irradiance exceeds the breakdown threshold. Kojima et al. [20] and Magnotti et al. [12,21] used a stretcher with multiple optical ring cavities to avoid the breakdown. Furthermore, the laser pulse was broadened after passing through the stretcher, which means a larger camera gate width can be used, and thus more signals can be collected. However, the optical ring cavities are usually very complex, expensive, and hard to align. Egermann et al. [7] used a telescope consisting of two cylindrical lenses and a 1000 mm focal-length spherical lens to avoid the breakdown by enlarging the focal volume. Dreyer et al. [6] also used a 1000 mm focal-length spherical lens to let the focused laser beam pass through the measurement volume to guarantee the breakdown occurred outside the measurement volume. However, the 1000 mm focal-length spherical lens will lower the spatial resolution due to the larger beam waist. It is a trade-off between Raman scattering intensity and spatial resolution.

Another approach is to reflect the laser beam into the measurement volume again by prisms/mirrors [22–26] or by a special optical cavity, so-called cavity-enhanced Raman spectroscopy (CERS) [27]. Other methods like coherent anti-Stokes Raman spectroscopy (CARS) or resonance Raman spectroscopy can also highlight the Raman scattering of the specific molecule [1,3]. Nevertheless, those kinds of approaches are either hard to implement or expensive.

In this paper, we proposed a short-gated SRS method using nanosecond 355-nm laser pulses, which is relatively low-cost and easy-to-implemented. Several new strategies have been utilized to separate SRS from interference signals and enhance the SNR. The proposed SRS was applied in a counterflow diffusion flame and validated by the numerical simulation.

## 2. Experimental method

### 2.1 Burner and flames

In-situ laser diagnostics of temperature and major species were performed on a counterflow diffusion flame with a strong soot particle formation. Details of the counterflow burner can be found in our recent studies [28–30]. In brief, ethylene and nitrogen were mixed and fed through the lower nozzle. The $N_2$ was used to dilute the $C_2H_4$. The flow rate of $N_2$ and $C_2H_4$ is 3.0926 and 3.0882 standard liter per minute (SLM), respectively. The air was introduced from the upper nozzle with a flow rate of 7.0583 SLM as the oxidizer. The co-flow of $N_2$ was used to stabilize the flame and isolate the experiment volume from ambient air. In our recent work [30], The soot volume fractions in the counterflow diffusion flame were measured by LII, which was calibrated by laser extinction measurements. The soot particles were generated from a height above the burner (HAB) of approx. 9.5 mm near the flame reaction zone and transported towards the stagnation plane at the HAB of 7.5 mm. The maximum soot volume fraction is approx. 0.1 ppm.

### 2.2 Laser diagnostic setup

The schematic of the laser diagnostic setup is presented in Fig. 1. A frequency-tripled Q-switch Nd: YAG laser (Brilliant b, Quantel) with the wavelength of 355 nm operating at the repetition rate of 10 Hz and pulse duration of approx. 5 ns was employed as the excitation source. The full laser power of approx. 125 mJ/pulse was chosen to maximize the SRS signal intensity. The polarization of the laser beam was horizontal at the output of the laser head and then was rotated to the vertical direction by an image rotator before focusing into the flame by a plano-convex UV spherical lens with a focal length of 300 mm. On the spectra measurement side, two plano-convex UV spherical lenses were used to collect the SRS signal into the slit of the spectrometer (ISOPLANE SCT 320, Princeton Instruments) after passing through an image rotator and a polarizer filter. The UV polarizer was set in the horizontal direction (along the x-axis) to maximize the Raman scattering. The grating was chosen to be 1800 groove/mm, and the slit was set to 150 μm, which resulted in a spectral resolution of 15.23 cm$^{-1}$. The signal was then split into spectra and then imaged onto an ICCD camera (PI-MAX 4, Princeton Instruments). The spatial resolution of the whole spectra collection system was measured to be 10.557 μm/pixel.

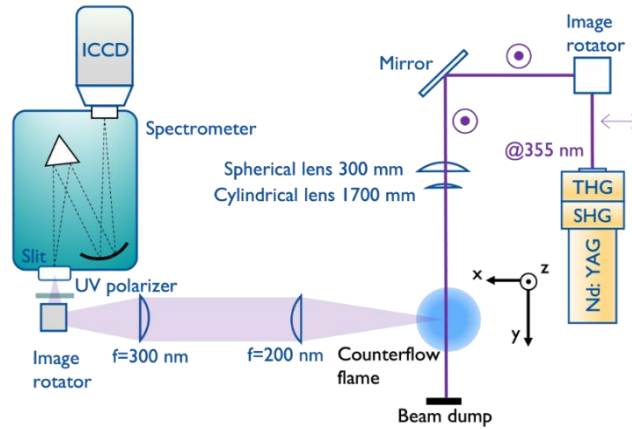

**Fig. 1** Schematic of the laser diagnostic setup

To avoid laser-induced breakdown occurring at the full laser power, an additional cylindrical lens with a focal length of 1700 mm was placed behind the 300-mm spherical lens on the laser beam path. Figure 2 shows the three-dimensional sketch of the focusing laser beam profile, and the beam profiles at five different positions were recorded qualitatively by a beam profiler with a resolution of 36 μm/pixel (LBS-100, Spiricon). The laser beam was focused in the horizontal direction (x-axis) by the cylindrical lens and was not changed in the vertical direction (z-axis). The beam diameter in the vertical direction was controlled by the 300-mm spherical lens and measured to be 0.13 mm by the ICCD camera, while the beam width in the horizontal direction was measured to be 1.2 mm by the beam profiler. The SRS signal was integrated along the slit for 10.8 mm. Thus the measurement volume was 1.2×10.8×0.13 mm$^3$ (x-y-z). The spatial resolution of 0.13 mm in the vertical direction (z-axis) is relatively high compared to the spatial resolution in the direction of the x- and y-axis. Because the measurements were carried out in a laminar counterflow diffusion flame with its spatial distribution uniform in the x-y plane, a low spatial resolution in the x-y plane was acceptable.

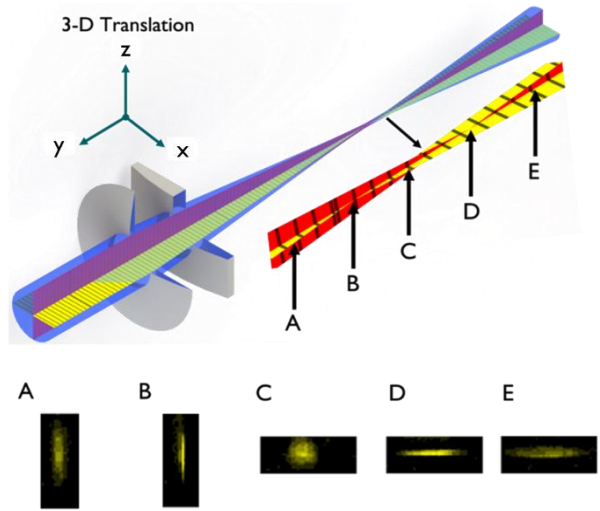

**Fig. 2** Three-dimensional sketch of the extended focal volume of the laser beam. A-E are beam profiles of 5 points along the laser beam (y-axis). B and D are the focal points of the cylindrical lens and spherical lens, respectively. The focal point D is the center of the measurement volume

## 3. Result and discussion

### 3.1 Temporal evolution of spectroscopy

To minimize the interference from the laser interacting with soot particles during SRS measurement, the ICCD gate time was accurately controlled in this work. Time-resolved spectra at different delay times were collected at the HAB of 9.5 mm, as depicted in Fig. 3. The starting time of the ICCD gate was set to sequentially shift from 80 ns to 95 ns with the interval of 1.5 ns, while the gate width time was kept at 5 ns. In total, 11 frames were collected and the spectra were averaged 50 times for every frame. The Stokes N$_2$ Raman signals were identified at around 386 nm with the Raman shift wavenumber of 2330 cm$^{-1}$. Its profile was fitted to measure temperatures. The Raman spectra mainly interfered with three sources of signals, CH-LIF signals in the $B^2\Sigma^- - X^2\Pi$ transition at 2363 cm$^{-1}$, CN-LIF in the $B^2\Sigma^+ - X^2\Sigma^+$ transition at 2275 cm$^{-1}$ [31], and a broadband spectrum.

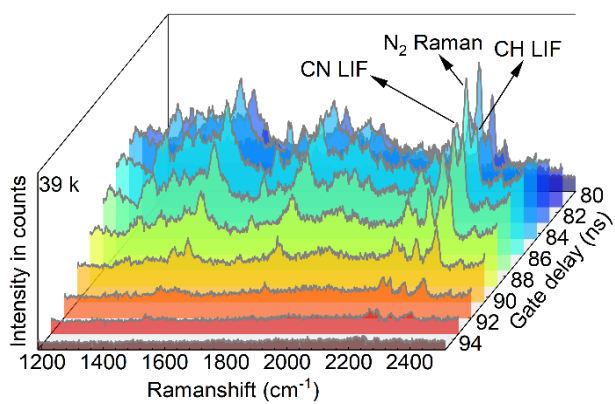

**Fig. 3** Time-resolved Raman spectra including LIF as the interference. The spectra were taken at 80-95 ns gate delay with HAB 9.5 mm. The interval of gate delay is 1.5 ns. Each spectrum is an average of 50 acquisitions.

The Raman scattering of N$_2$ @ 2330 cm$^{-1}$ appears at 81.5 ns delay with the broadband background, while the CH and CN LIF spectra have not been observed yet. Then, the N$_2$ Raman signal intensity grew up rapidly in the first 3 frames from 81.5 ns to 84.5 ns before getting diminished and vanished after 92 ns. The CH and CN LIF signals became dominant over the Raman spectra after 87.5 ns. After the delay time of 95 ns, both LIF and Raman spectra are too weak to be observable.

Figure 4 (a) further demonstrated the signal intensities of these spectra as a function of the delay time. The CH LIF @ 2363 cm$^{-1}$, N$_2$ Raman scattering @ 2330 cm$^{-1}$, and broadband emission averaged intensity in the range of 1670 cm$^{-1}$ – 1690 cm$^{-1}$ were considered. The signal intensity for each point in Fig. 4 (a) can be viewed as an integral of the signal temporal profile over the 5-ns gate time, as indicated by the blue arrays. The instantaneous Rayleigh scattering was used as the time flag of the laser pulse. The time-resolved Rayleigh scattering was measured using a sequentially moved 5-ns ICCD gate and fitted by a convolution function between a Gaussian distribution, i.e. laser pulse profile, and a square function distribution function, i.e. the gate time. Figure 4 (b) shows the calculated laser pulse profile. The maximum point of the laser intensity profile occurred at the ICCD delay time of approx. 87.5 ns and thus was denoted as the time of 0. Figure 4 (c) summarized the SNR of the Raman and LIF spectra at different delay times. As shown in Fig. 4 (a), the Raman signal reaches maximum earlier than LIF because the Raman scattering can be seen as a spontaneous process, while the LIF occurs in nanoseconds due to the complete absorption, excitation, collision, and return to the ground state [3]. Thus, the gate delay time can be chosen appropriately to maximize the SNR of the Raman and minimize that of LIF. In comparison, the broadband emission shows a rapid increase and slow decay. A possible reason is that the broadband emission may consist of broadband Raman and LIF signals of PAH after laser ablation of soot particles. When the laser intensity increases, the PAHs Raman signal dominates the broadband emission, and for the decay, PAHs LIF dominates. It is deduced that the LII signals of soot may not be significant here, because the LII signals usually occur between 400 nm and 800 nm and have a longer delay time than PAHs LIF [32,33]. Based on the above analysis, the collection time centered at -2 ns corresponding to the ICCD gate time of 83-88 ns can maximize the SNR of Raman and thus be chosen to be the ICCD delay time in the following measurement.

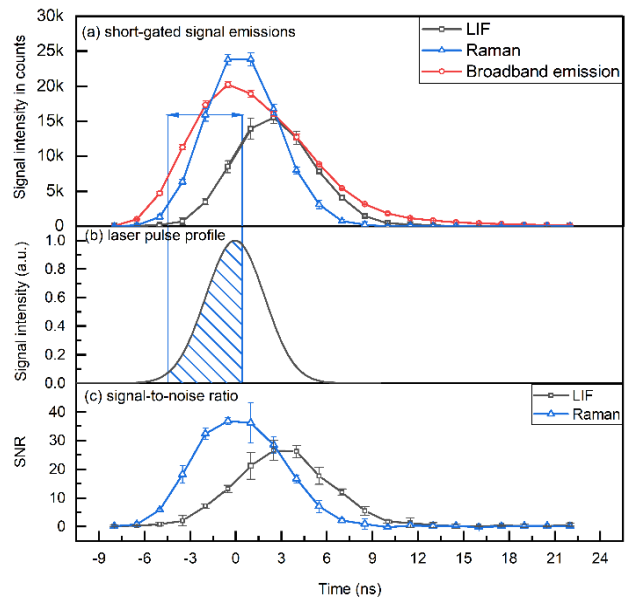

**Fig. 4** (a) Temporal profile of Raman scattering, LIF, and broadband emissions; (b)

Laser pulse profile by fitting with Gaussian function; (c) Signal-Noise-Ratio of Raman scattering and LIF.

**3.2 Nonlinear Raman spectra**

The collected Raman signal was then compared directly with the simulated Raman spectrum to fit the temperature. Based on the theory of quantum mechanics, the Raman signal is induced by the energy exchange from the inelastic collisions between molecules and photons. The modified vibrational and rotational energy of molecules is then reflected in the shifted radiation frequency. Depending on whether the molecules gain or lose energy, Stokes Raman scattering ($v_s = v_0 - \Delta\tilde{v}$) and Anti-Stokes Raman scattering ($v_{as} = v_0 + \Delta\tilde{v}$) occurs. The size of the shifted wavenumber is a function of the energy level of the molecule:

$$\Delta\tilde{v} = \tilde{G}(v') - \tilde{G}(v'') + \tilde{F}(J') - \tilde{F}(J'') \tag{1}$$

The superscript ' indicates the final energy state, the superscript " indicates the initial energy state, $v$ is the vibrational quantum number, and $J$ is the rotational quantum number. For the diatomic molecules, the vibrational term $\tilde{G}(v)$ and rotational term $\tilde{F}(J)$ can be described as

$$\tilde{G}(v) = \tilde{v}_e(v + 1/2) - x_e\tilde{v}_e(v + 1/2)^2 \tag{2}$$

$$\tilde{F}(J) = \left(B_e - \alpha_e(v + 1/2)\right)J(J + 1) - \left(D_e - \beta_e(v + 1/2)\right)J^2(J + 1)^2 \tag{3}$$

Here the $\tilde{v}_e$, $x_e$, $B_e$, $\alpha_e$, $D_e$, and $\beta_e$ are molecule specific constants [1,34,35]. The intensity for individual Stokes Raman transitions with a collection solid angle at 90° can be calculated by:

$$I_{v',J'} = \frac{\alpha_{zz}^2 \pi^2 (v_0 - \Delta\tilde{v})^4}{\epsilon_0^2} \cdot N \cdot p_{v'',J''} \cdot I_{laser} \tag{4}$$

where $\alpha_{zz}$ is the polarizability matrix element of vertically polarized scattering induced by vertically polarized laser; $v_0$ is the wavenumber of laser; $\epsilon_0$ is the permittivity of free space; $N_i$ is the number density of molecule $i$; $p_{v'',J''}$ is the population distribution of the initial energy state with $v'',J''$, which is prescribed by the Boltzmann distribution; and $I_{laser}$ is laser irradiance [3]. According to the Placzek polarizability theory and selection rules of diatomic molecules, the polarizability matrix element $\alpha_{zz}$ can be expressed as:

When $\Delta v = +1, \Delta J = 0$, $\quad \overline{(\alpha_{zz})^2} = (v+1)\dfrac{h}{8c\pi^2 \Delta\tilde{v}}\left[(\alpha')^2 + \dfrac{4}{45}b_{J',J''}(\gamma')^2\right]$ (5)

When $\Delta v = +1, \Delta J = \pm 2$, $\quad \overline{(\alpha_{zz})^2} = (v+1)\dfrac{h}{8c\pi^2 \Delta\tilde{v}}\left[\dfrac{4}{45}b_{J',J''}(\gamma')^2\right]$ (6)

where $c$ is the speed of light, $\alpha'$ and $\gamma'$ are the mean and anisotropy invariants of the derived polarizability tensor [36], $b_{J',J''}$ is the Placzek-Teller coefficients [3].

The population distribution $p_{v,J}$ as well as $\alpha_{zz}$ determine the line intensity for each vibration-rotational transition together at the wavenumber shift of $\Delta v$. By accumulating each line with broadening in a Voight shape, the Stokes-Raman profile at $\Delta v = 1$ can be then simulated and utilized to fit the experimental spectra. The fitting process involves the baseline fitting of the broadband spectrum and the nonlinear fitting of the Raman spectra which have been subtracted from the baseline. The Raman spectra fitting utilized a non-linear fitting function 'lsqnonlin' that is based on the algorithm of trust-region-reflective algorithm, which is integrated within Matlab.

Figure 5 demonstrates the baseline fitting (red line) and $N_2$ Raman spectrum fitting (blue line) of the flame at the HAB of 6.94 mm, 8.49 mm, and 11.07 mm. These three positions correspond to the low-temperature region, the sooting flame region, and the high-temperature region, respectively. For 95% confidence intervals, the uncertainties caused by the nonlinear fitting procedure were around ±9 K for the low-temperature region, ±2 K for the high-temperature region, and ±3 K for the sooting region.

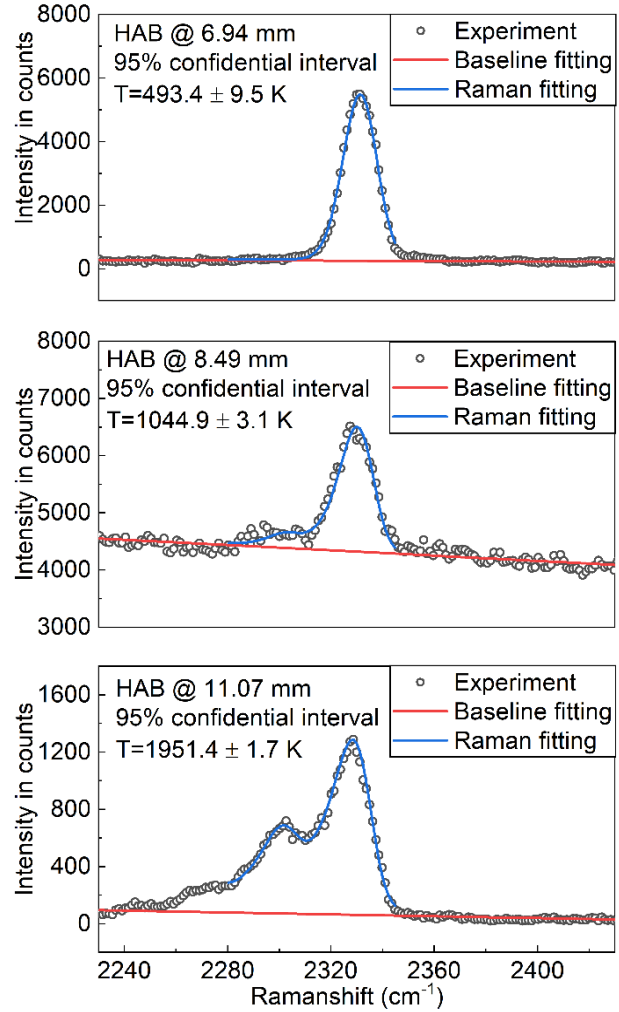

**Fig. 5** Baseline fitting (red line) and $N_2$ Raman spectrum fitting (blue line) of flame at (a) HAB 6.94 mm, (b) HAB 8.49 mm, and (c) HAB 11.07 mm.

The total intensity of the Stokes Raman transitions was utilized to obtain the mole fraction of major species. According to a theoretical analysis by Schrotter and Klockner [37], the total intensity of the vibration-rotation band can be simplified as that of the pure vibrational transition averaged over all orientations of the molecule [3]. This fact indicates that the temperature influences the integrated Raman scattering intensity $I_{tot}$ in two ways: (1) $I_{tot}$ is inversely proportional to the vibrational partition function $\left(1 - e^{-hc\tilde{v}_e/kT}\right)$; (2) $I_{tot}$ is proportional to the number density $N_i$ of species $i$ which is then inversely proportional to the temperature according to the ideal gas law. Therefore, at a given temperature $T$, the total intensity of Raman scattering $I_{tot}$ of species $i$ can be calibrated with a reference condition with the temperature of $T_0$ and the mole fraction

of $x_0$ to obtain the mole fraction of $x_i$ according to

$$x_i = \frac{I_i}{I_0}\frac{T_i}{T_0}\frac{\left(1-e^{-hc\tilde{v}_e/kT_i}\right)}{\left(1-e^{-hc\tilde{v}_e/kT_0}\right)}x_0 \quad (7)$$

**3.3 Demonstration in the counterflow diffusion flame**

The short-gated SRS method was then utilized in the measurement of temperature and species of the ethylene-based flames with soot particles formed in the flame. Figure 6 demonstrates the measured temperature profile of the flame at each HABs with its simulated temperature profile by FlameMaster [38]. The measurement precision is evaluated by the standard deviation of 5 measurements. The deviation reaches as high as approx. ±70 K at the sooting area due to the interference. In addition, it should be noted that the deviation in the low-temperature region is generally higher than that at the high-temperature region because the sensitivity of the Stokes-Raman spectral shape on the temperature at low temperatures is lower than that at high temperatures. In high-temperature non-sooting regions, the standard deviation of the measured temperature is as low as 50 K, while it reaches up to 130 K for low temperatures. The accuracy of the measurement is evaluated by comparing it with the simulation. In general, the measurement temperature profile matches well with the simulation. More details about the simulation can be found in our recent work [30]. The highest temperature can be measured with accuracy within 50 K. The measured temperature profile in the sooting area agrees well with the simulated temperature profile.

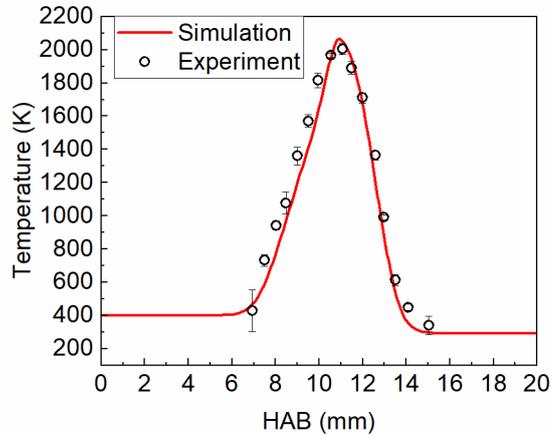

**Fig. 6** Measured and simulated temperature profile of flame.

Figure 7 further demonstrates the typical Raman spectra of $O_2$, $C_2H_4$, and $N_2$ at different HABs in the counterflow burner after subtracting the baseline of these spectra. The baseline was obtained by fitting the original spectra using 2-degree polynomial curves. The $O_2$ spectra at the Raman shift of 1550 cm$^{-1}$, $C_2H_4$ v3 spectra at 1343 cm$^{-1}$, and the $N_2$ spectra at 2331 cm$^{-1}$ were integrated and processed to evaluate the mole fraction of $O_2$, $C_2H_4$, and $N_2$, respectively. The $C_2H_4$ v3 is chosen considering its stronger intensity than the $C_2H_4$ v2 signal and its spectral location farther away from $O_2$. For the species of $N_2$ and $O_2$, the reference point is set at the HAB of 15 mm with $x_{N2}$ of 0.79 and $x_{O2}$ of 0.21, while for the species of $C_2H_4$, the reference point is set at the HAB of 7 mm with the $x_{C2H4}$ of 0.4997.

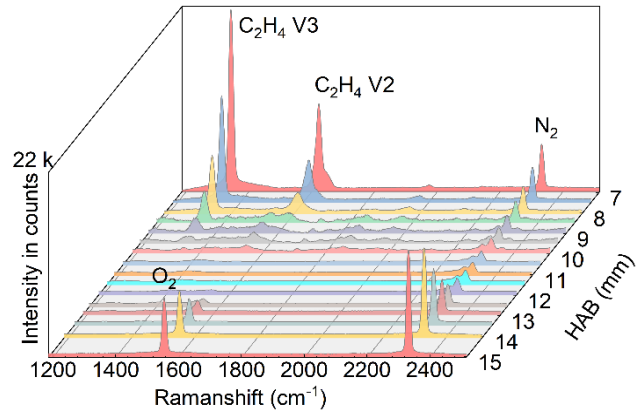

**Fig. 7** Raman scattering of different species at different HABs after subtracting the baseline

Figure 8 shows the measured mole fraction of $N_2$, $O_2$, and $C_2H_4$. The measurement uncertainty is derived according to Eq. 7 based on the uncertainties of the total Raman signal intensity $I_{tot}$ and that of the measured temperatures $T$, which are estimated by their respective standard deviations. The measured values match well with the simulation results, indicating a general good accuracy. A large uncertainty occurs at the boundaries with low temperatures, which is caused by the large deviation of the temperature measurement during the Raman spectra fitting.

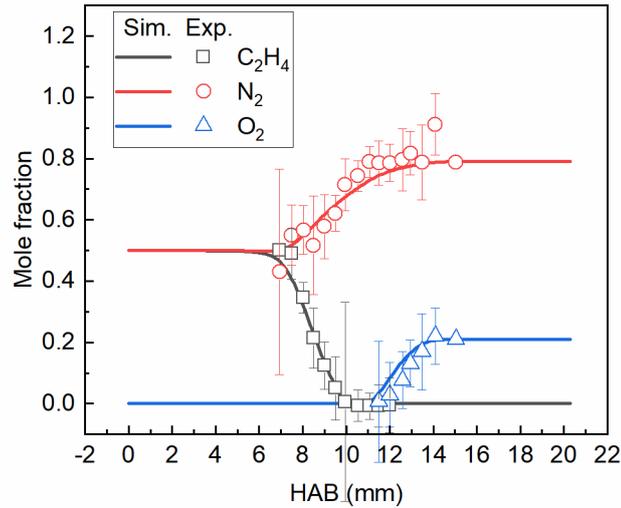

**Fig. 8** The comparison of measured with simulated mole fraction of $N_2$, $O_2$, and $C_2H_4$

## 4. Conclusion

This study demonstrates a short-gated spontaneous Raman spectroscopy in a fuel-rich combustion environment (soot volume fraction ~ 0.1 ppm). The optical setup with spherical and cylindrical lenses avoids the laser-induced breakdown and guarantees a high laser intensity. The short gate was elaborately controlled to suppress the LIF and LII signals. The temperature was measured by fitting the Raman scattering with simulated spectra of $N_2$, based on which the mole fraction can be further evaluated according to the ideal gas law. To validate the short-gated Raman spectroscopy, the temperature and mole fraction of ethylene-based counterflow diffusion flame are measured. The temperature and mole fraction agrees well with the simulated temperature profile and calculated mole fraction, even in the sooting area, indicating a good accuracy of the new thermometry method. Compared to all the other methods that have been used, this new method does not require complex equipment (such as a picosecond laser source/ICCD or Kerr gate) and achieves the temperature measurements without calibration, showing the potential for in-situ laser diagnostics of other particle-laden flames in an easy-to-implement and low-cost way.

## Acknowledgement

The work was mainly funded by the Deutsche Forschungsge-meinschaft (DFG,